**Design of Face Centered Cubic Co$_{81.8}$Si$_{9.1}$B$_{9.1}$ with High Magnetocrystalline Anisotropy**

*Dr. Jiliang Zhang, Dr. Guangcun Shan\*, Dr. Yuefei Zhang, Dr. Fazhu Ding and Prof. Dr. Chan Hung Shek\**

Dr. Jiliang Zhang, Dr. Guangcun Shan\*, Prof. Dr. Chan Hung Shek

[a]Department of Materials Science, City University of Hong Kong, Hong Kong
Email:gcshan-2@my.cityu.edu.hk, apchshek@cityu.edu.hk

Dr. Guangcun Shan,

[b]School of Instrument Science and Opto-electronic Engineering, Beihang University, Beijing 100191, China;

Dr. Yuefei Zhang

[c]Institute of Microstructure and Property of Advanced Materials, Beijing University of Technology, Beijing 100124, China

Dr. Fazhu Ding

[d]Institute of Electrical Engineering, Chinese Academy of Sciences, Beijing 100190, China
[e]University of Chinese Academy of Sciences, Beijing 100049, China

Despite the composition close to glassy forming alloys, face centered cubic (FCC) Co$_{81.8}$Si$_{9.1}$B$_{9.1}$, designed based on Co9B atomic cluster (polyhedral), are synthesized as single-phase ribbons successfully. These ribbons, with grain sizes of ca. 92 nm, show supreme ductility and strong orientation along (111), which couples with shape anisotropy leading to high magnetocrystalline anisotropy comparable to Co rich Co-Pt nanoscale thin films, with a coercivity of 430 Oe and squareness of 0.82 at room temperature. The stability and magnetic behaviors of the phase are discussed based on experimental electronic structure. This work not only develops low cost Co-based materials for hard magnetic applications, but also extends the atomic cluster model developed for amorphous alloys into the design of new crystalline materials.

Metallic cobalt (Co) has been widely studied as an important magnetic material and its magnetic properties can vary significantly depending on its structures:[1] the face centered cubic (FCC, stable above 450 °C) phase shows good soft magnetic behaviours, and amorphous CoSiB materials based on the structure (with a general composition at ca. $Co_{80}(Si,B)_{20}$) are one of best soft magnetic materials, used in many high-performance electric devices;[2] the hexagonal close-packed (HCP, stable below 450 °C) Co shows hard magnetic behavior because of the structural anisotropy, and its compounds with rare earth (*RE*) elements (e.g. $SmCo_5$) or noble metals (e.g. CoPt) do show significantly enhanced hard magnetic behaviours due to the high anisotropy of RE elements or strong $3d$-$5d$ hybridizations.[3,4] The high cost of these materials inhibits their applications as permanent magnets in many cases. On another hand, many of these compounds show unsatisfactory mechanical performance. The fabrication of nanomaterials especially nanowires does improve the hard magnetic behaviours significantly, [5, 6] but the complex process and oxidization are main drawbacks.

In metallic glasses, it is generally agreed that there are one or two kinds of atomic clusters (short-range orders) dominating the structure and thus controlling the physical properties. [7] It is reported that based on different dominant clusters, CoSiB metallic glasses with close compositions but significantly different structure and magnetic behaviors (from ferromagnetism to anitferromagnetism) can be achieved.[8, 9] In infinite solid state solutions, it is also found that their atomic structures are not completely disordered. Instead, dominant atomic clusters determining the magnetic properties also exist in these compounds, and can vary depending on the compositions. [5,6,10] Thus, it is possible to design this kind of magnetic compounds based on the dominant atomic clusters by using the atomic cluster model previously developed for amorphous alloys.[8]

CoB/CoSiB metallic glasses are widely studied in both structure and magnetic properties since 1970s. The atomic clusters favouring glassy formation have to be avoided, because the absence of anisotropy in glassy states will deteriorate the hard magnetism. According to Co-B phase diagram,[1] there is a Co rich FCC phase, $Co_{23}B_6$, stable at high temperature. Thus the cluster favouring the FCC structure can be extracted from this compound,[11] as $Co_9B$ polyhedral (see the inset structure in **Figure 1a**). In order to reduce the cost and increase the magnetic anisotropy, it is better to include Si and B as much as possible while remaining the FCC or HCP Co-type structure. The maximum solubility of Si is 10 at. % (Co:Si=9:1) in HCP Co and 18 at. % in FCC Co.[1] Si concentration more than the maximum solubility may favour the amorphous state instead of Co structure. Thus the composition is designed as $Co_9BSi$, normalized as $Co_{81.8}Si_{9.1}B_{9.1}$. The synthesized structure is expected to be FCC Co type, because the dominant cluster is from FCC $Co_{23}B_6$. As a result, in our work, we obtained the FCC Co-type CoSiB ribbons successfully, with high magnetocrystalline anisotropy. The structural stability and magnetic behaviours of the compound can also be understood based on experimental electronic structures.

The X-ray diffraction (XRD) pattern in **Figure 1a** shows the dominant (111) reflection of the FCC phase with no visible trace of amorphous hump, affirming the single-phase nature of the as-made ribbons. The absence of other reflections suggests the strong orientation along (111). The morphologies of these ribbons indicate the good ductility of these samples, as shown in the inset of Figure 1(a). The lattice parameter estimated from the (111) reflection is $a$=3.514 Å, smaller than the value 3.537-3.558 Å of FCC Co at room temperature.[12] It is in agreement with the small atomic size of the substituting elements B and Si.

The (111) reflection peak is broadened obviously, suggestive of the nanocrystalline nature. The apparent crystallite size was estimated as ca. 92 nm using the Scherrer's formula.[13] The nanocrystals of $Co_{81.8}Si_{9.1}B_{9.1}$ were confirmed by transmission electron microscopy (TEM), as shown in **Figure 1b**. All observed crystals are less than 100 nm in critical size, in agreement with XRD results. No diffuse ring observed in selected area electron diffraction (see the inset in **Figure 1b**) confirms the absence of any amorphous phases. Most crystals are not spherical but elongated along the spinning direction. The isotropic shape surely benefits the hard magnetic behaviours.

Magnetic hysteresis loops were measured under applied fields along longitude direction and perpendicular to ribbons at room temperature. As shown in **Figure 2**, $Co_{81.8}Si_{9.1}B_{9.1}$ exhibits high magnetocrystalline anisotropy with (111) easy direction. Such strong anisotropy was mainly observed in nanoscale thin film or nanomaterials of anisotropic shapes. [6, 14]

The magnetization is almost saturated at 2000 Oe along the easy direction close to the field to saturate soft CoB/CoSiB amorphous alloys, and at 10000Oe along the perpendicular direction close to the field for bulk and nanoscale Co in both FCC and HCP structures. However, the saturation magnetization ($M_s$) is only 31.7 emu/g, corresponding to 28 emu per Co gram, much smaller than the reported $M_s$ in bulk Co and most nanoscale Co.[15, 16] Obviously the reduced $M_s$ is not due to the dilution effect of the substituted elements. It was reported that Co nanoparticles of 2 nm in size show a smaller $M_s$, very likely owing to the surface oxidation. [17] Apparently, it is not the case here. Though antiferromagnetic CoSiB amorphous alloys are also reported, the small filed for saturation along (111) rules out the possibility. Thus the reduced $M_s$ should result from the partial delocalization of Co 3$d$ electrons due to the strong interaction with B or Si.

The coercivity ($H_c$) is ca. 430 Oe, in both easy and hard directions, as evident in the inset of **Figure 3**. The coercivity is much larger than bulk Co and most of reported nanoscale HCP and FCC Cobalt (10-300 Oe),[15, 17] and close to some Co-Pt thin films.[14] Despite the reduced

Co magnetic moment, the saturation remanence ($M_r$) is still up to 26 emu/g with a high squareness ($M_r/M_s$) of 0.82. With proper tuning, the materials will be promising hard magnetic materials.

To reveal the origin of the improvement in hard magnetization, magnetic domains in the samples were observed by magnetic force microscaopy (MFM). The samples show (see **Figure 3**) distinctive feature of dark and bright stripe magnetic domian structure. The stripe domains are often observed in thin ferromagnetic films and platelets with weak perpendicular or oblique anisotropy.[18] In ferromagnetic films, the period of strip domains is comparable to the film thickness. [18] However, the period (width) of the strip domain in the $Co_{81.8}Si_{9.1}B_{9.1}$ ribbon is ∼ 90 nm, which is much less than the thickness of the ribbon (∼ 20 μm). Instead, the period is very close to the longitudal size of the nanocrystal grains. In the view, the stripe magnetic domains in the $Co_{81.8}Si_{9.1}B_{9.1}$ ribbons here should mainly result from the anisotropic shape of CoSiB nanocrystals. Thus the hard magnetic behaviour of the materials can be further improved by tuning grain sizes.

Despite the fact that the composition of the FCC $Co_{81.8}Si_{9.1}B_{9.1}$ is very close to compositions of many CoSiB amorphous alloys, neither amorphous nor crystalline impurities were observed in XRD patterns and TEM observations. It is already reported that two CoSiB amorphous alloys having very close compositions show distinctive differences in their magnetisms, due to different short-range orders or clusters. [8, 9] To understand the origin of the special structure and magnetism of the present $Co_{81.8}Si_{9.1}B_{9.1}$ phase, electronic structures around fermi level ware measured on the present $Co_{81.8}Si_{9.1}B_{9.1}$ phase, HCP Co and a soft magnetic CoSiB alloy $Co_{63.1}B_{27}Si_{9.9}$. Because the major phase after crystallization of soft magnetic CoSiB amorphous alloy is generally the FCC Co structure, the electronic structure of soft magnetic CoSiB amorphous alloy should be similar to that of FCC Co.[8, 19]

Compared with the electronic structures of HCP Co (see **Figure 4**), many soft magnetic CoSiB alloys show similar feature around the fermi level: peak electronic density of state (DOS, N(E)) at fermi level followed by a shoulder at 1.5 eV (black arrows in **Figure 4**). [8, 20] Both HCP Co and CoSiB amorphous alloy have large DOS close to the peak value at their fermi level, while the FCC $Co_{81.8}Si_{9.1}B_{9.1}$ exhibits a small DOS at the fermi level. It is evident that the FCC $Co_{81.8}Si_{9.1}B_{9.1}$ is more stable than both hcp Co and amorphous CoSiB alloys. As the electronic structure around the Fermi level determines the magnetic behaviours, the Co atoms in both HCP Co and amorphous CoSiB show similar magnetic moment, in agreement with literatures. The shoulder at 1.5 eV corresponds to the electrons in the spin-down states, while the peak at Fermi level is from spin-up electrons.[20] For the FCC $Co_{81.8}Si_{9.1}B_{9.1}$, the shift of peak towards high binding energy and the absence of the shoulder indicate that the strong interactions between Co and B/Si leads to the delocalisation of Co 3$d$ electrons and thus low Co moments. It is also supported by the fact that the hallmark peak of localized B 2$p$-Co 3$d$ hybridisation at ~ 8 eV in soft magnetic CoSiB is not observed in the present FCC $Co_{81.8}Si_{9.1}B_{9.1}$ phase.[21] Such delocalization yielding significantly reduced magnetic moments was also observed in FeBY glassy alloys.[22] Obviously, the compositions of FCC CoSiB phase are tuneable to achieve a balance between better magnetic performance and structural stability.

Based on the Co9B cluster extracted from FCC $Co_{23}B_6$, the low-cost FCC $Co_{81.8}Si_{9.1}B_{9.1}$ ribbon was successfully produced. Despite the high content of metalloid Si and B, the as – made ribbons are ductile and show significantly enhanced hard magnetic behaviours similar to Co-Pt nanoscale thin film, with a room-temperature coercivity of 430 Oe, much larger than the value of most reported bulk and nanoscale Co. The wall thickness of magnetic domains is close to the critical size of $Co_{81.8}Si_{9.1}B_{9.1}$ nanocrystals, indicating the important role of these nanocrystals on the magnetism. The structural stability and special magnetism of the present materials are well explained based on the measured electronic structures. The results also show that the hard magnetic behaviours can be further improved without destruction of the structure, although a balance between structural stability and magnetic performance has to be taken into account.

**Experimental Section**

*Synthesis of FCC $Co_{81.8}Si_{9.1}B_{9.1}$*: the mixtures (10 g) of pure constituent elements Co chunks (99.99 wt. %), Si chunks (99.99 wt. %), and B particles (99.9 wt. %) under a Ti-gettered argon atmosphere. The alloy ingots were remelted four times to improve compositional homogeneity. Using these master ingots, ribbon samples with thickness of ~30 μm and width of ~2 mm were produced by a single roller melting-spinning apparatus at a wheel surface velocity of 40 m/s.

*Characterization and measurements*: Phase identification was conducted on a Philips X'pert x-rays diffractometer (Cu Kα, 0.15406 nm). Microstructure observations were done with a high-resolution transmission electron microscope (HRTEM; model No. TECNAI 20 G2, FEI Company, Hillsboro, OR) operated at 200 kV. Ribbons for TEM observations were thinned by twin-jet electropolishing using a $HClO_4$-$C_2H_6O$ solution (volume ratio 1:8) at ~ −30 °C. These specimens were finally cleaned by low-angle ion milling for no more than 5 min for TEM observation. Magnetic measurements were carried out using a Cryogenic vibrating sample magnetometer with a maximum field of up to 20000 Oe. Surface morphologies and magnetic domain structures of ribbons were observed using a Vecco atomic force microscope and a magnetic force microscope in the tapping/lift scanning mode with a CoCr-coated tip magnetized downward. Measurements of electronic structure were performed using x-ray photoelectron spectroscopy with a monochromatized Al Kα radiation (1486.6 eV) in an ULVAC-PHI 5802 system (Kanagawa, Japan).

**Acknowledgements**
The work was financially supported by National key research and development program (no. 2016YFC1402504) by MOST of China and funded by a General Research Fund from Research Grants Council, Hong Kong SAR Government (Project number: CityU 11211114).

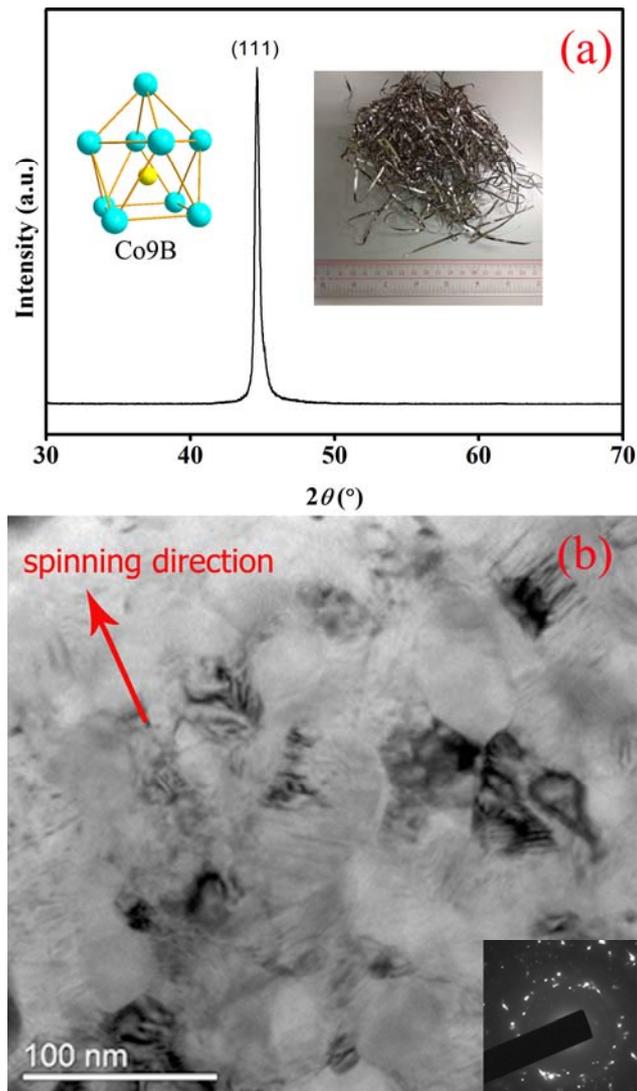

**Figure 1.** PXRD pattern (a) and TEM observation (b) of as-cast $Co_{81.8}Si_{9.1}B_{9.1}$ ribbons. The inset in (a) shows pictures of ribbons and the Co9B cluster extracted from $Co_{23}B_6$ (yellow balls for B and blue balls for Co). The inset in (b) gives the selected area electron diffraction corresponding to the observed zone.

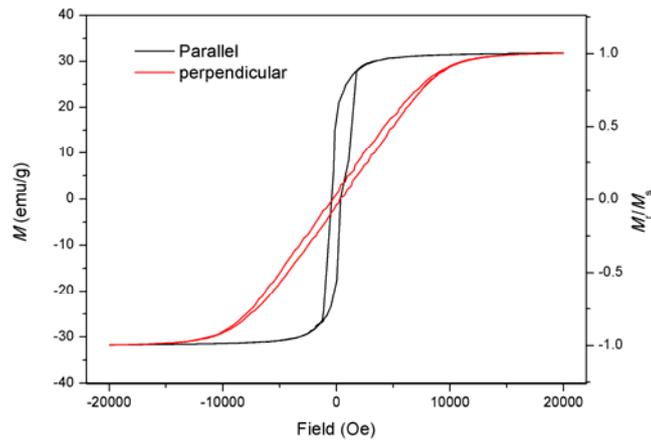

**Figure 2.** Hysteresis loops of $Co_{81.8}Si_{9.1}B_{9.1}$ ribbons measured with a field of 20000 Oe applied along (black) the longitudinal direction and perpendicular (red) to ribbons.

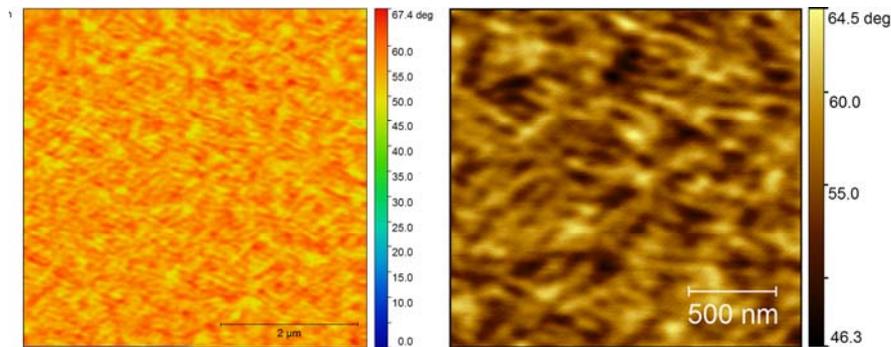

Figure 3

**Figure 3.** Magnetic domain structure (left, square: 5x5 µm) of the $Co_{81.8}Si_{9.1}B_{9.1}$ ribbon and its magnified views (right, square: 2x2 µm).

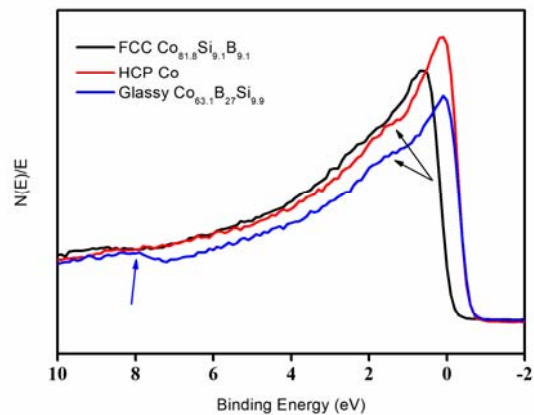

**Figure 4.** Electronic structures of the $Co_{81.8}Si_{9.1}B_{9.1}$ phase(black), bulk HCP Co (red) and a typical soft magnetic CoSiB glassy alloy ($Co_{63.1}B_{27}Si_{9.9}$). The black arrows mark a shoulder next to the peak electronic DOS, and the blue arrow indicates the shoulder induced by B-Co interactions.

**Title**
**Design of Face Centered Cubic Co$_{81.8}$Si$_{9.1}$B$_{9.1}$ with High Magnetocrystalline Anisotropy**

ToC figure

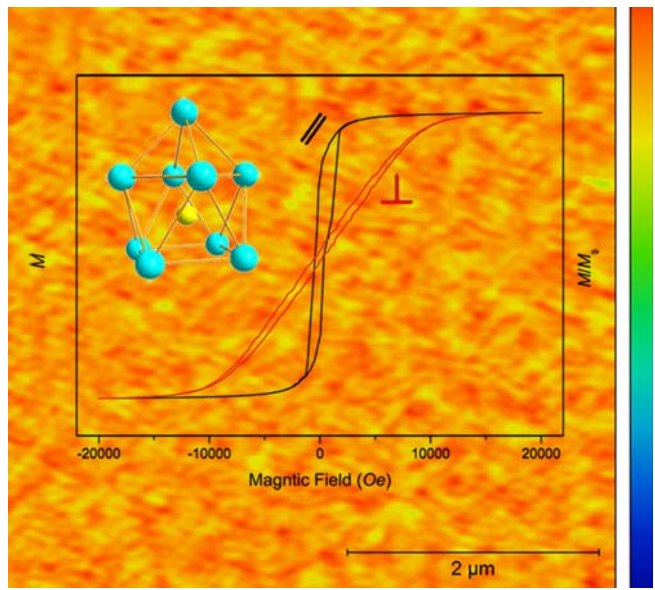